\documentclass[a4paper]{aa}
\usepackage{graphicx}

\def\degmark{^\circ}
\def \rsun {\ifmmode$R$_{\odot}\else R$_{\odot}$\fi}
\def \nh {N${\rm _H}$}
\def \hcm {\hbox {\ifmmode $ atom cm$^{-2}\else atom cm$^{-2}$\fi}}
\def \src {Her\,X-1}
\def\approxgt{\mathrel{\hbox{\rlap{\lower.55ex \hbox {$\sim$}}
        \kern-.3em \raise.4ex \hbox{$>$}}}}
\def\approxlt{\mathrel{\hbox{\rlap{\lower.55ex \hbox {$\sim$}}
        \kern-.3em \raise.4ex \hbox{$<$}}}}
\newcommand{\mc}{\multicolumn}

\newcommand {\einstein} {{\it Einstein}}

\newcommand {\degree} {$^{\circ}$}

\newcommand {\rchisq} {$\chi_{\nu} ^{2}$}

\begin{document}

\title{BeppoSAX spectroscopy of the Hercules X-1 short-on state}

\author{T.~Oosterbroek\inst{1} \and A.N.~Parmar\inst{1} \and D. Dal
Fiume\inst{2} \and M. Orlandini\inst{2} \and A. Santangelo\inst{3}
\and S. Del Sordo\inst{3} \and A.~Segreto\inst{3}}

\offprints{T. Oosterbroek (toosterb@astro. estec.esa.nl)}
\institute{Astrophysics Division, Space Science Department of ESA, 
ESTEC, P.O. Box 299, NL-2200 AG Noordwijk, The Netherlands
\and Istituto TESRE, CNR, via Gobetti 101, I-40129 Bologna, Italy
\and IFCAI, CNR, via La Malfa 153, I-90146 Palermo, Italy}

\thesaurus{(02.01.2; 08.09.2 Her X-1; 08.14.1; 13.25.5)}

\date{Received 15 July 1999/ Accepted 11 October 1999}

\authorrunning{T. Oosterbroek et al.}

\maketitle 



\begin{abstract}
We present results of a 5.7~day duration BeppoSAX observation 
of the short-on state of \src\ and a  
short observation during the decline of the preceding main-on state.
The 0.1--10~keV spectra can be fit with
a power-law and blackbody model together with Fe 
emission features at 1.0~keV and 6.5~keV.
During the later stages of the short-on state there are long intervals
when the absorption is $\approxgt$$5 \times 10^{22}$~atom~cm$^{-2}$.
These intervals become longer and occur $\sim$0.3~day earlier in each orbital
cycle as the short-on state progresses. During the intervals of
high absorption the 0.1~keV blackbody is still clearly detected. 
This may indicate the presence of separate scattered and absorbed spectral
components, although other explanations such as 
partial covering or a partially ionized absorber cannot be excluded.
During the rest of the short-on state the ratio of flux in the 
blackbody compared to the power-law is consistent with that in the
main-on state.
This supports the view that much
of the 35~day modulation is caused by an energy
independent process, such as electron scattering.
The discovery of strong absorption late in the short-on state is
consistent 
with the predictions of the warped disk model (Petterson \cite{pe:77},
see also Schandl \&
Meyer \cite{s:94}) where the end of the short-on state is caused
by the accretion disk moving into the line of sight to the
neutron star. 
The pulse phase difference between the blackbody
and the power-law maxima is $250 \pm 20\degmark$ in both 
observations (separated by 0.43 of a 35~day cycle).
This constant phase difference is consistent with the blackbody 
originating at the inner edge of a precessing accretion disk.

\keywords{accretion, accretion disks -- Stars: individual: \src\
-- Stars: neutron -- X-rays: stars} \\  

\end{abstract}

\section{Introduction}

\src\ is an eclipsing X-ray pulsar with a pulse period of
1.24\,s and an orbital period of 1.70~days (Tananbaum et al.\
\cite{t:72}; Giacconi et al. \cite{g:73}). The source exhibits a
35~day X-ray intensity cycle consisting of a $\sim$10 day duration main
on-state between $\Phi _{35} = 0.0-0.31$ (using the
ephemeris of Scott \& Leahy (\cite{s:99}) where $\Phi _{35} = 0.0$
is defined as the turn-on to the main-on state)
and a fainter $\sim$5 day duration secondary, or short, on-state
between $\Phi _{35}$ = 0.57--0.79.
At other $\Phi _{35}$
\src\ is still visible as a low-level X-ray source (Jones \& Forman
\cite{j:76}). This modulation has been ascribed to a tilted precessing
accretion disk (Gerend \& Boynton \cite{g:76}) and an accretion disk
corona Schandl \& Meyer (\cite{s:94})
that periodically obscure 
the line of sight to the
neutron star .
In the warped disk model of Schandl \& Meyer (\cite{s:94})
and Schandl (\cite{s:96}),
the onset of the main-on state is caused by the edge of the accretion
disk moving out of the line of sight to the neutron star and the end of 
the main-on state is caused by the corona crossing the line of sight.
The onset of the main-on state is associated with an increase
in photoelectric absorption by cold material, while no such increase is
seen during its more gradual decay 
(e.g., Becker et al. \cite{b:77}; Parmar et al. \cite{p:80}).
In addition, a regular pattern of narrow X-ray intensity dips are observed
which repeat every 1.65~days. These are probably caused by obscuration
from cold material in the outer regions of the accretion disk
(Crosa \& Boynton \cite{c:80})

The broad-band on-state X-ray spectrum of \src\ is known to be complex
and consists of at least the following components: (1) a power-law
with a photon index, $\alpha$, of $\sim$0.9 in the energy
range 2--20~keV. (2) Cyclotron absorption (Tr\"umper et
al. \cite{t:78}; Mihara et al. \cite{m:90}) at energies $\approxgt$20~keV.  
(3) A broad Fe emission feature near 6.4~keV (Pravdo et
al. \cite{p:77}; Choi et al.\ \cite{c:94}). (4) A 
blackbody with a temperature, kT, of $\sim$0.1~keV
(Shulman et al. \cite{s:75}; Catura \& Acton \cite{c:75};
McCray et al. \cite{m:82}; Mavromatakis \cite{m:93}; Vrtilek
et al. \cite{v:94}; Choi et al. \cite{c:97}; Oosterbroek et al.\
\cite{o:97}, hereafter O97) and (5) a broad emission feature between
0.8--1.4~keV, which is probably unresolved Fe~L shell emission (McCray et
al. \cite{m:82}; Mihara \& Soong \cite{m:94}; O97).

Pulse-phase spectroscopy using the \einstein\ Objective Grating (OGS)
and Solid State Spectrometers (SSS) in the energy range 0.15--4.5~keV
by McCray et al. (\cite{m:82}) revealed that the maximum intensities of
the blackbody and power-law components are shifted by
240$\degmark$ during the main-on state. 
It is likely that the blackbody component results from
hard X-rays that are reprocessed in the illuminated
inner regions of the accretion
disk.  McCray et al. (\cite{m:82}) note that the phase of maximum
intensity of the unresolved 0.8--1.4~keV feature appears coincident
with that of the blackbody.  The pulse-phase dependence of the
blackbody component was investigated with better quality data by O97
who confirmed the above phase-difference, and established that the Fe-L 
line and the blackbody exhibit a similar phase-dependent behavior 
during the main-on state.

In this {\it paper} we report on a BeppoSAX observation 
designed to investigate the nature of the 
short-on state modulation. 
An additional short observation was performed during the declining phase of the
preceding main-on state for comparison purposes. 
We concentrate on the 0.1--10 keV spectrum of \src, which
contains all the continuum and reprocessed components. 

\section{Observations}
\label{sec:observations}

\begin{figure}
\centerline{\includegraphics[height=8cm,angle=-90]{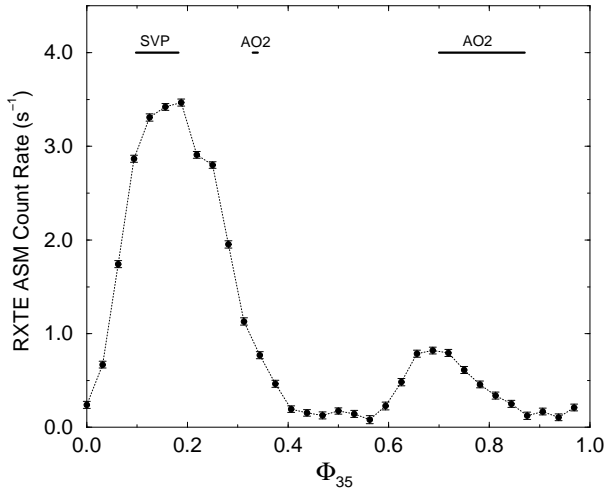}}
\caption[]{The mean 35 day intensity cycle of \src\ measured by the
           RXTE ASM. The phases, $\Phi _{35}$, of the Science
           Verification Phase (SVP) and AO2 main- and short-on 
           state observations (AO2) are indicated}
\label{fig:phase_long}
\end{figure}

Results from the Low-Energy Concentrator Spectrometer (LECS;
0.1--10~keV; Parmar et al. \cite{p:97}), and the Medium-Energy Concentrator
Spectrometer (MECS; 1.8--10~keV; Boella et al. \cite{b:97}) 
on-board BeppoSAX are presented.
The MECS consists of two grazing incidence
telescopes with imaging gas scintillation proportional counters in
their focal planes. The LECS uses an identical concentrator system as
the MECS, but utilizes an ultra-thin entrance window and
a driftless configuration to extend the low-energy response to
0.1~keV.

The main-on state observation was performed during the trailing edge of the
on-state, significantly later in $\Phi _{35}$ than the observation
reported in O97. The 5.7~day duration short-on state
observation covered most of the short-on state (and possibly
some low-state at the end), but did not include the turn-on. 
The main-on state observation started at 1998
June~27 08:34 and lasted until June~28 02:56 (UTC).
The short-on state observation lasted from 1998 July~9 20:38 until
July~15 15:13 (UTC). Fig.\ \ref{fig:phase_long} illustrates the
$\Phi _{35}$ coverage of the three BeppoSAX observations of \src\ 
discussed here.  The dashed line shows the mean
35~day 2--12~keV intensity profile obtained by folding the Rossi X-ray
Timing Explorer (RXTE) All-sky Monitor (ASM) data obtained between
1996 February~20 and 1999 February~24 over the best-fit period of 34.75~days.
The difference in $\Phi _{35}$ between the mid-times of the two
observations discussed here is 0.43.

LECS and MECS data were
extracted centered on the \src\ position using radii of 8\arcmin\ and 
4\arcmin, respectively. 
Good data were selected from intervals when the minimum
elevation angle above the Earth's limb was $>$5\degree\ ($>$4\degree\
for the LECS) and when the
instrument configuration was nominal using the SAXDAS 1.8.0 data
analysis package.  Spectral analysis was performed with the response
matrix from the 1997 September release of SAXDAS for the MECS and a
response matrix appropriate for the source position in the LECS
field of view using LEMAT 3.5.6. The total main-on state
exposures are 18~ks and 25~ks for the LECS and MECS, respectively.
The total short-on state
exposure times are 125~ks and 165~ks for the LECS and MECS,
respectively.  

\begin{figure*}
\centerline{
   \hbox{\includegraphics[width=12cm,angle=-90]{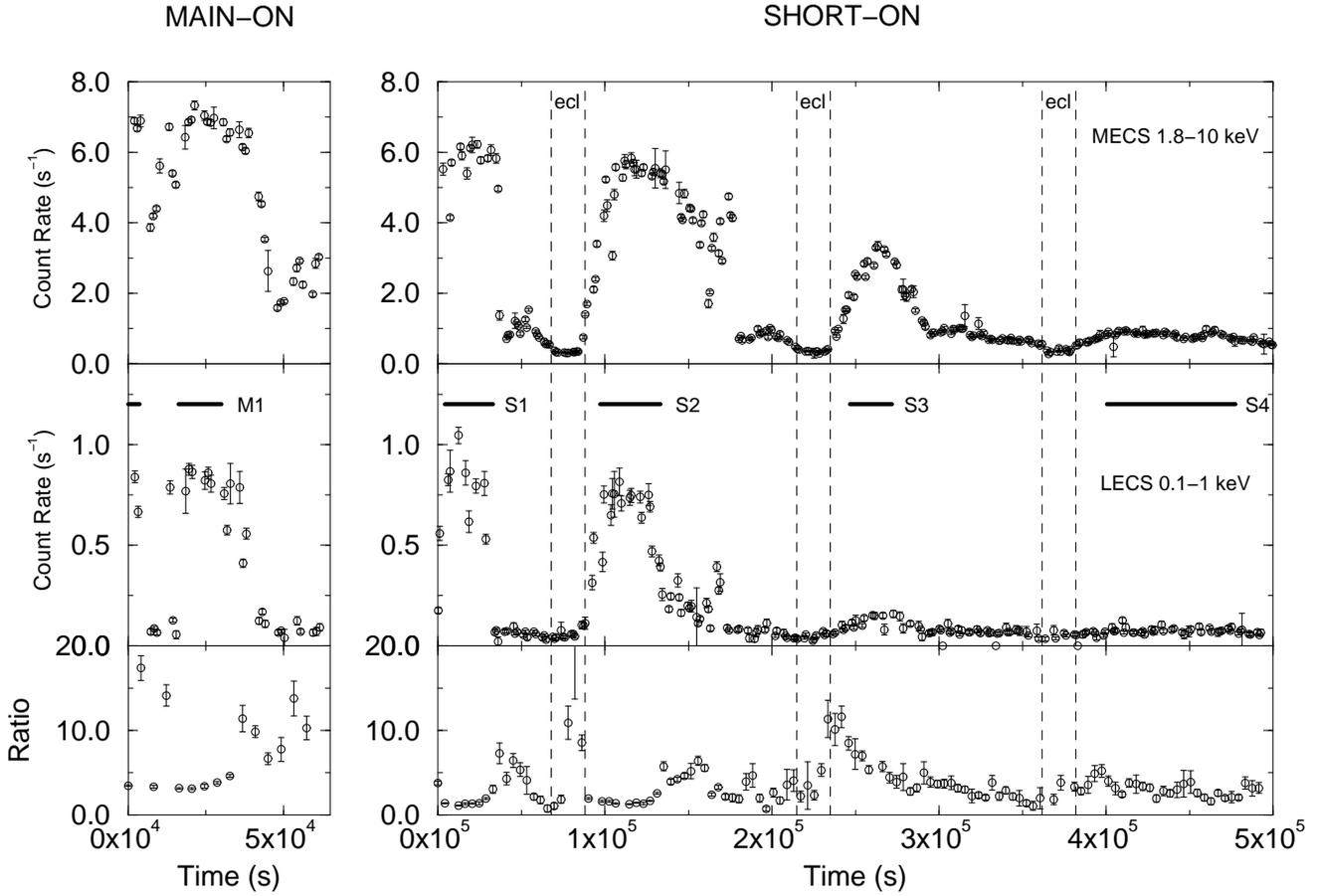}}}
\caption[]{MECS 1.8--10~keV (upper panels) and LECS 0.1--1.0~keV 
         (middle panels) lightcurves  obtained during the main- (left) and
         short-on state (right) observations with a binning of 1024~s. Note
         the different temporal scales. 
         The predicted eclipse times and durations, using the
         ephemeris of Deeter et al. (\cite{d:91}), are indicated with
         dashed lines. The intervals M1 and S1 to S4, used to extract 
         continuum spectra, are indicated by horizontal bars (see
         Sect.~\ref{subsec:spec}). Times are
         seconds since the start of each observation. The lower panels
         show the LECS hardness ratio (4.0--10~keV/0.1--1.0~keV) plotted
         with a binning of 4096~s}
\label{fig:lc}
\end{figure*}

\section{Analysis and results}
\label{sec:analysis}

\subsection{X-ray lightcurve}
\label{subsec:lc}

The 0.1--1.0~keV LECS and 1.8--10~keV MECS lightcurves of both observations 
are shown in Fig.~\ref{fig:lc} together with the hardness ratios
(LECS counts between 4--10~keV divided by those between
0.1--1.0~keV). The ordinate extrema are the same for both on-states, but
the time axes have different scales, with the main-on state plots expanded
by about a factor of 2. Eclipse intervals based on the orbital ephemeris
of Deeter et al. (\cite{d:91}) and assuming an on-state eclipse duration
of 5.5~hour are indicated.

A number of new and interesting features are evident.
Soon after the start of the main-on state observation there is 
an interval of dipping activity
(the modulation is much stronger in the LECS than the MECS, consistent
with the known energy dependence of dipping), followed
by a dip-free interval and a longer interval of deep dipping at the
end of the observation.
The short-on state observation covers parts of 4 orbital cycles and
includes 3 eclipse intervals. Both the LECS and the MECS show
a gradual reduction in count rate, with this effect being more
pronounced in the LECS, such that the fourth orbital cycle appears
to be absent, whereas a small modulation is still 
visible in the MECS. 
Superposed on this decay are the eclipses
and what is normally taken to be dipping activity. However,
this appears to be
present for up to 20~hrs during each orbital cycle, whereas
the main-on state dip duration is usually 5--10~hr
(see e.g., Reynolds \& Parmar \cite{r:95};
Scott \& Leahy \cite{s:99}).
During the first three orbital cycles, the centroid of the emission occurs
at successively earlier orbital phases, while for the fourth orbital
cycle no strong variation within the cycle is observed. 
This is also seen in  the 
hardness ratio
plot which shows intervals of increased hardness (consistent with
increased absorption) that occur progressively earlier in each
of the first 3 (and possibly the fourth) orbital cycles.
The peaks of the MECS centroids and the increases in LECS
hardness ratio seen in Fig.~\ref{fig:lc} are separated by an
average of $\sim$1.4~days. This means that the intervals of
strong absorption march back by $\sim$7~hr each orbital cycle.
This rapid marching back is in strong contrast to the main-on state
dips which have a period only 0.5~hr less than the orbital one
(Scott \& Leahy \cite{s:99}).
There may also be a narrow intensity dip in the MECS lightcurve,
similar to those seen in the on-state, towards the end of
the second cycle of the short-on state.

\begin{table*}
\begin{center}
\caption[]{Spectral fit parameters and 90\% confidence
uncertainties for the phase-averaged spectral fits. 
The spectra have been fit with the ``standard '' model.
The energies of the narrow Fe-L lines were
fixed at the best-fit ASCA values in Mihara \& Soong
(\cite{m:94}). The equivalent blackbody radius
assumes a distance of 6.6 kpc (Reynolds et al. \cite{r:97}).
The (0.1--10~keV) intensity is for all spectral components
}
\begin{tabular}{llllll}
\hline
\noalign {\smallskip}
Parameter              & \mc{5}{c}{Spectrum} \\
                       &  \hfil M1  &  \hfil S1 & \hfil S2 & \hfil S3
                       & \hfil S4 \\
\hline
\noalign {\smallskip}
LECS/MECS exposure (ks)& 6.6/8.6 & 4.5/7.3 & 7.4/10.9 & 7.6/9.7 & 19.8/26.6 \\
$\alpha$               & $0.753 \pm ^{0.017} _{0.012}$ 
                       & $0.787 \pm ^{0.021} _{0.011}$
                       & $0.802 \pm ^{0.020} _{0.009}$
                       & $0.53$
                       & $0.50  \pm 0.03$ \\
Blackbody kT (keV)     & $0.097 \pm ^{0.006} _{0.002}$ 
                       & $0.094 \pm ^{0.004} _{0.003}$
                       & $0.094 \pm ^{0.002} _{0.005}$
                       & $0.12$
                       & $0.101 \pm 0.008$ \\
Equiv. BB radius (km)  & $215 \pm 16$
                       & $209 \pm 13$
                       & $202 \pm 10$
                       & $262$ 
                       & $67 \pm ^{59} _{67}$ \\
\nh\ $(\times 10^{19}$ atom cm$^{-2}$) & $0.15 \pm ^{0.10} _{0.15}$ 
                       & $0.19 \pm ^{0.13} _{0.11}$ 
                       & $0.22 \pm  0.07$
                       & $750$
                       & $<$0.6 \\
Fe-L 0.91 keV EW (eV)  & $100 \pm 30$
                       & $95 \pm ^{30} _{20}$
                       & $70 \pm 30$
                       & $0.0$
                       & $4 \pm ^{41} _{4}$ \\
Fe-L 1.06 keV EW (eV)  & $65 \pm 30$ 
                       & $60 \pm 30$
                       & $40 \pm 40$
                       & $0.0$
                       & $125 \pm 50$ \\
Fe-K line energy (keV) & $6.445 \pm ^{0.13} _{0.03}$ 
                       & $6.50  \pm 0.07$
                       & $6.52 \pm 0.05$
                       & $6.156$
                       & $6.55 \pm 0.05$ \\
Fe-K line FWHM (keV)   & $0.83 \pm ^{0.26} _{0.13}$ 
                       & $0.92 \pm ^{0.12} _{0.08}$
                       & $0.84 \pm 0.14$ 
                       & 2.94
                       & $0.80 \pm ^{0.14} _{0.09}$\\
Fe-K line EW (eV)      & $305 \pm 50$ 
                       & $470 \pm ^{200} _{75}$
                       & $405 \pm 70$
                       & $1600$
                       & $1080\pm ^{100} _{180}$ \\
Intensity (erg cm$^{-2}$ s$^{-1}$) & $8.0\times 10^{-10}$ 
                       & $6.6\times 10^{-10}$
                       & $6.0\times 10^{-10}$
                       & $2.6\times 10^{-10}$
                       & $6.4\times 10^{-11}$ \\
$\chi ^2$/dof          & 104.2/89 & 94.2/89 & 131.9/89 & 383/83 & 85.6/84 \\
\noalign {\smallskip}                       
\hline
\label{tab:ave_fits}
\end{tabular}
\end{center}
\end{table*}

\subsection{X-ray spectrum}
\label{subsec:spec}

In order to study the spectral evolution during the parts of the 35~day
cycle observed here, 
5 phase averaged spectra were extracted. The selected intervals are
indicated in Fig.~\ref{fig:lc} and are labelled as M1 
and S1 to S4. Interval M1 covers a dip-free part of the main-on
state, while S1 and S2 cover the peaks of the first two
short-on orbital cycles and have a similar overall intensity to M1.
Interval S3 covers part of the third short-on orbital cycle, where the LECS
intensity is severely reduced, while S4 covers most of the final short-on
cycle, which may also include low-state emission.

The spectra were rebinned to oversample the full
width half maximum (FWHM) of the energy resolution by
a factor 3 and to have additionally a minimum of 20 counts 
per bin to allow use of the $\chi^2$ statistic. 
Additionally a 2\% systematic error was added to the
uncertainties.
Events were selected in the energy ranges
0.1--4.0~keV (LECS) and 1.8--10~keV (MECS) 
where the instrument responses are well determined and sufficient
counts obtained. 
The photoelectric absorption
cross sections of Morisson \& McCammon (\cite{m:83}) and the
solar abundances of Anders \& Grevesse (\cite{a:89}) are used throughout.
A factor was included in the spectral fitting to allow for normalization 
uncertainties between the instruments. This was constrained
to be within the usual range of 0.8--1.0 (LECS/MECS) during all fitting. 

\begin{figure*}
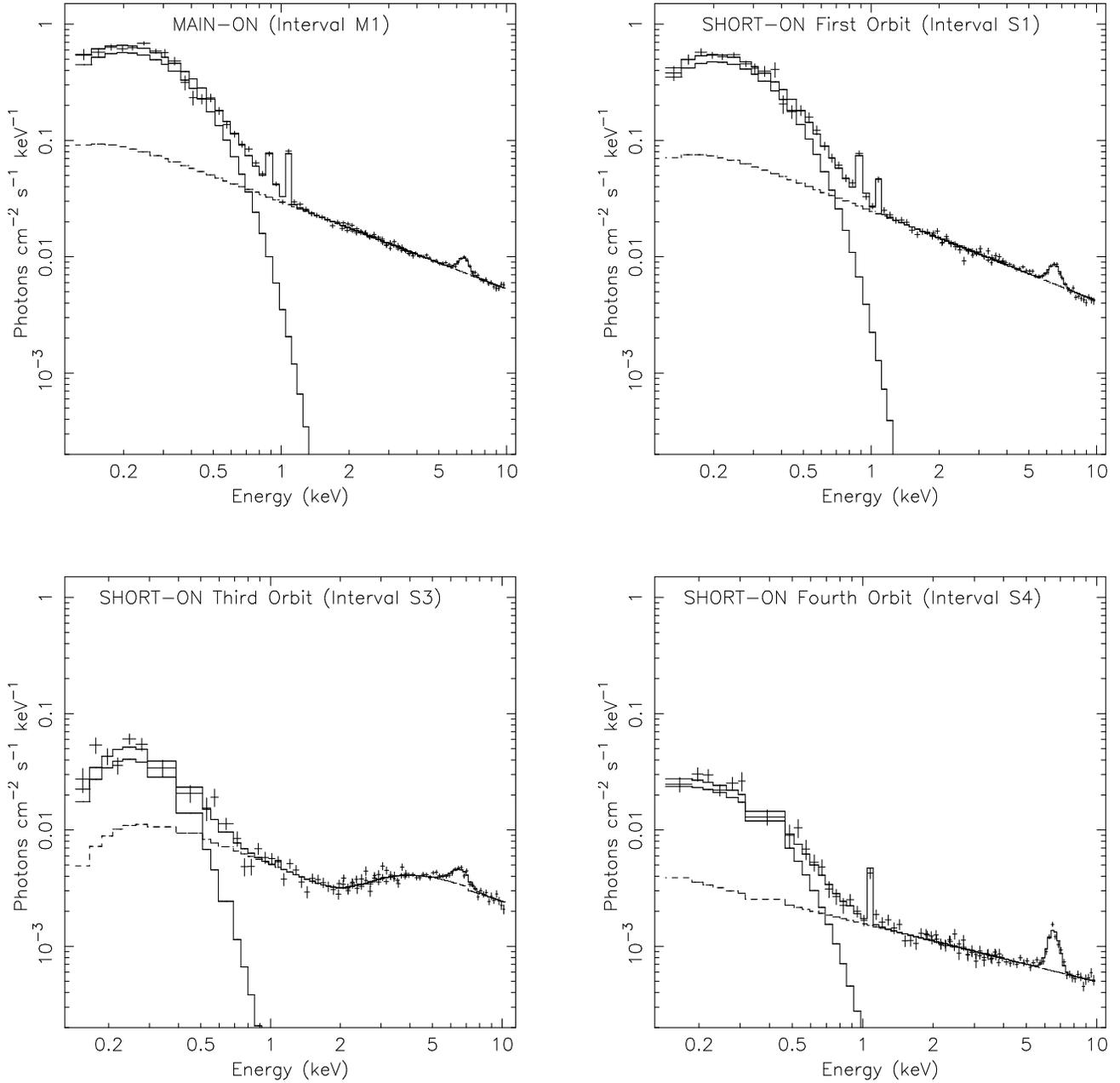

\mbox{ 
\hspace{0.0cm}
\includegraphics[height=8.0cm,angle=-90]{m1_uf.ps}
\hspace{1.0cm}
\includegraphics[height=8.0cm,angle=-90]{s1_uf.ps}}
\vspace{1.0cm}\\
\mbox{
\hspace{0.0cm}
\includegraphics[height=8.0cm,angle=-90]{s3_pcf_uf.ps}
\hspace{1.0cm}
\includegraphics[height=8.0cm,angle=-90]{s4_uf.ps}}
\caption[]{LECS and MECS \src\ spectra during the main-on 
(M1) and short-on states (S1, S3, and S4).
The solid lines show the unfolded spectrum obtained with the 
``standard'' spectral model (see Table~\ref{tab:ave_fits}), while for S3 a
partially covering absorber is used (see
Table~\ref{tab:s3_fits}). 
The contributions from the blackbody and power-law 
components are indicated separately.
The same scales have been used for all four panels.
The effect of the large amount of 
absorption required in the S3 spectral fits is clearly seen as  
a change in the spectral slope around 2~keV}
\label{fig:spectra}
\end{figure*}

\begin{table}
\begin{center}
\caption[]{Partial covering fit results to the S3 and S4 spectra.
The energies of the narrow Fe-L lines were
fixed at the best-fit ASCA values in Mihara \& Soong (\cite{m:94}).
f is the fraction of the flux that undergoes extra absorption, ${\rm N_{PCF}}$}
\begin{tabular}{lll}
\hline
\noalign {\smallskip}
Parameter              & \mc{2}{c}{Spectrum} \\
 &        \hfil S3 & \hfil S4 \\
\hline
\noalign {\smallskip}
$\alpha$               & $0.864 \pm ^{0.02} _{0.04}$ 
                       & $0.74  \pm 0.04$ \\
Blackbody kT (keV)     & $0.083 \pm ^{0.006} _{0.002}$ 
                       & $0.089 \pm 0.015$ \\
Equiv. BB radius (km)  & $146 \pm ^{32} _{17} $
                       & $145 \pm ^{11} _{21} $ \\
\nh\ ($\times 10^{19}$ atom cm$^{-2}$) & $2.0 \pm ^{4.1} _{2.0}$ 
                       & $11.0  \pm 3.5$ \\
f                      & $0.723 \pm ^{0.023} _{0.013}$ 
                       & $0.35 \pm 0.03$      \\
${\rm N_{PCF}}$ ($\times 10^{22}$ atom cm$^{-2}$) & $6.03 \pm 0.45$ 
                       & $21 \pm ^{12} _{6}$ \\
Fe-L 0.91 keV EW (eV)  & $<$35
                       & $<$36 \\
Fe-L 1.06 keV EW (eV)  & $<$40 
                       & $80 \pm 45$ \\
Fe-K line energy (keV) & $6.54 \pm 0.10$ 
                       & $6.56 \pm 0.13$ \\
Fe-K line FWHM (keV)   & $1.59 \pm 0.27 $ 
                       & $0.76 \pm 0.23 $  \\
Fe-K line EW (eV)      & $590 \pm 80$ 
                       & $900 \pm ^{80} _{130}$ \\
$\chi ^2$/dof          & 115.7/82 & 78.9/82 \\
\noalign {\smallskip}                       
\hline
\label{tab:s3_fits}
\end{tabular}
\end{center}
\end{table}

Initially, all 5 spectra were fit with the absorbed
power-law and blackbody continuum
together with 2 broad Gaussian emission features at $\sim$1~keV
and $\sim$6.4~keV as used by O97. The ASCA 
Solid-State Imaging Spectrometer (SIS)
results of 
Mihara \& Soong (\cite{m:94}) indicate that the $\sim$1~keV
feature may be better modeled as two narrow lines at 
at 0.91~keV and 1.06~keV, at least in the low-state.
Since the full-width half maximum (FWHM) energy resolution of the
LECS is 0.2~keV at 1~keV, these features are 
not well resolved in the LECS and so the broad feature was replaced 
by two narrow lines with the energies fixed at the above
values in all subsequent fits.  
We refer to this as the ``standard'' model.
The fit results presented in
Table~\ref{tab:ave_fits} indicate that this model adequately
describes the M1, S1,
S2 and S4 spectra, but not the S3 spectrum where
a $\chi^{2}$ of 383 for 83 degrees of freedom (dof) is obtained. 
The fit quality of the S2 spectrum is somewhat
worse than to the M1, S1, and S4
spectra. This may be because of unresolved variability which was not
excluded from the accumulation (see the upper and middle panels
of Fig.~\ref{fig:lc}).
The S4 spectrum is significantly harder and the 
equivalent width (EW) of the Fe-K line is higher than 
the other spectra (Table~\ref{tab:ave_fits}). There is
no comparable change in the EW of the Fe-L lines.
The intense Fe-K feature
is clearly visible in the S4 spectrum shown in 
Fig.~\ref{fig:spectra}. There is no significant change in Fe-K line
energy between the spectra.
The equivalent blackbody radii are $\sim$200~km
for intervals M1, S1 and S2, decreasing to $\sim$150~km during S3 and
$\sim$70~km during S4.  The ratio of 0.1--10.0~keV flux in the blackbody
component compared to the powerlaw is 16\% for all the spectra except
S3, where is it 9\%. 

The effects of significant absorption are clearly seen in the S3 spectrum as a
change in spectral shape and increased curvature in the 
1--3 keV range (see Fig.~\ref{fig:spectra}). This implies an absorption
of $\approxgt$$10^{22}$~atom~cm$^{-2}$. 
However, substantial flux remains $\approxlt$0.5 keV, which should be 
{\it completely} absorbed with such a high absorption. There
are a number of possible explanations for such behavior: (1) 
the presence of separate ``scattered'' and ``absorbed'' spectral
components, (2) partial covering of the emitting region(s), and 
(3) absorption by partially ionized material 
such that the low $Z$ materials responsible for the majority of the 
absorption $\approxlt$0.5~keV are significantly
ionized, while the higher $Z$ elements are not. 
Possibilities (1) and (2) cannot be spectrally distinguished and
are referred to as ``partial covering'', although this should be taken
to include the possibility of separate scattering and absorbing regions.
Partial covering can be modeled using the 
{\tt pcfabs} model in {\sc xspec}.
Here a fraction, f, of the emission undergoes extra absorption,
${\rm N_{PCF}}$, while the rest is absorbed by a low value of
\nh, as before. 
The partial covering model gives an acceptable fit to the S3 
spectrum (a $\chi^2$ of 115.7 for 82 dof), while that of an ionized
absorber (the {\tt absori} model in {\sc xspec})
is somewhat worse ($\chi^2 = 133.9$ for 83 dof).
We therefore do not pursue further the 
ionized absorber fits since the 
partial covering model provides a better fit.

As previously stated, the ``standard model'' gives a significantly
harder spectrum ($\alpha = 0.50 \pm 0.03$) for interval S4 than for
the other spectra, where the average is $\sim$0.8. We have investigated
whether this apparent hardness may result from a large amount of
intrinsic absorption which, if not modeled correctly, results in an
anomalously hard spectral slope determination. The
partial covering model was also fit to the S4 spectrum, even though the
``standard'' model provides an adequate description of the spectrum ($\chi ^2 
= 85.6$ for 84 dof). This gives a partial covering fraction of
$0.35 \pm 0.03$ and extra absorption, ${\rm N_{PCF}}$, of
$(2.1 \pm ^{1.2} _{0.6}) \times 10^{23}$~atom~cm$^{-2}$. The best-fit value
of $\alpha$ is now $0.74 \pm 0.04$, much closer to the values obtained
in the other fits. This suggests that an explanation for the
hard spectra (relative to the peak of the main-on state 
where $\alpha \sim 0.9$) seen at times from \src\ is the effects of large
amounts of unresolved absorption, together with partial covering.

\begin{figure*}
\centerline{\mbox{
\includegraphics[width=8cm]{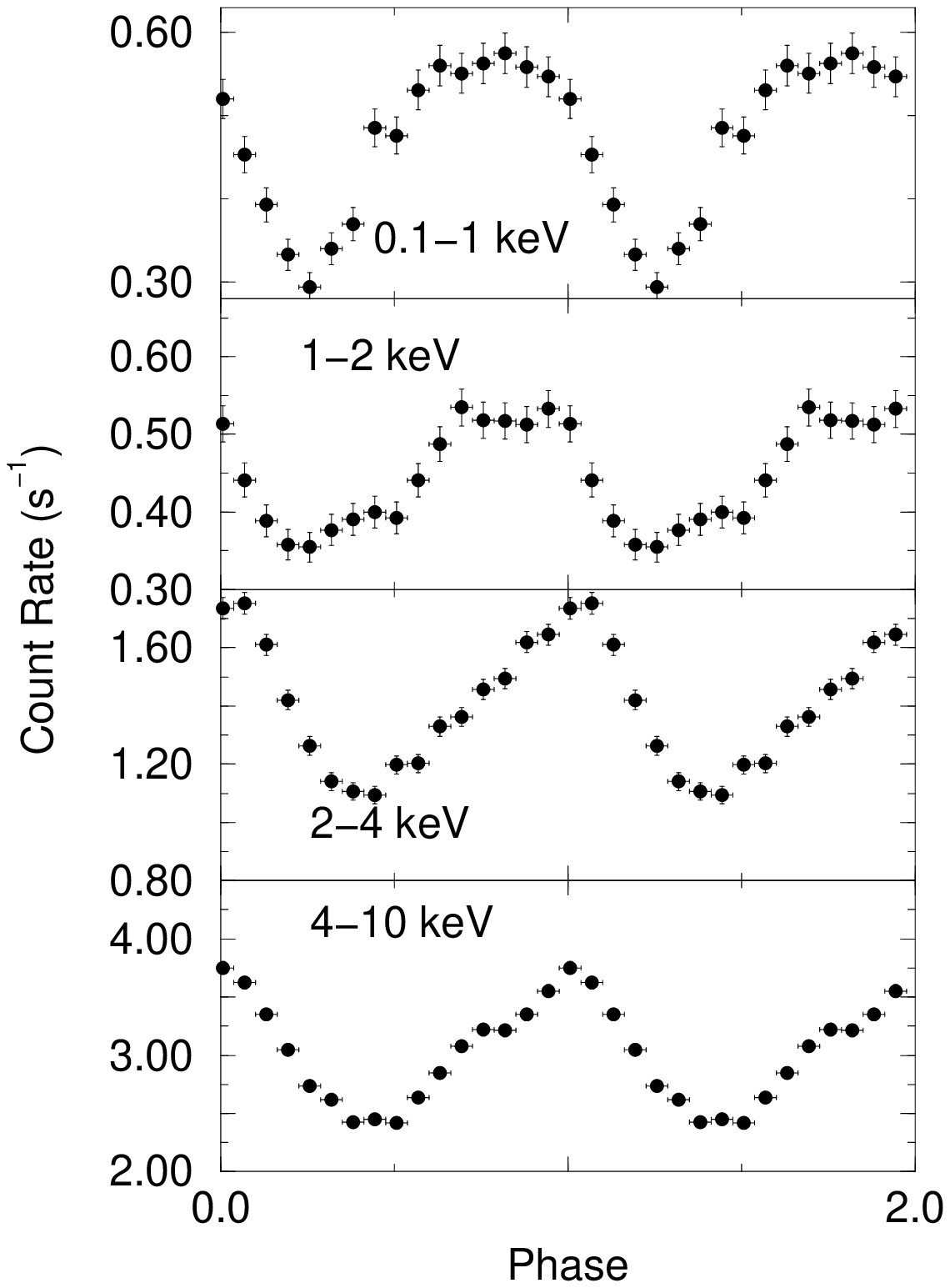}
\includegraphics[width=8cm]{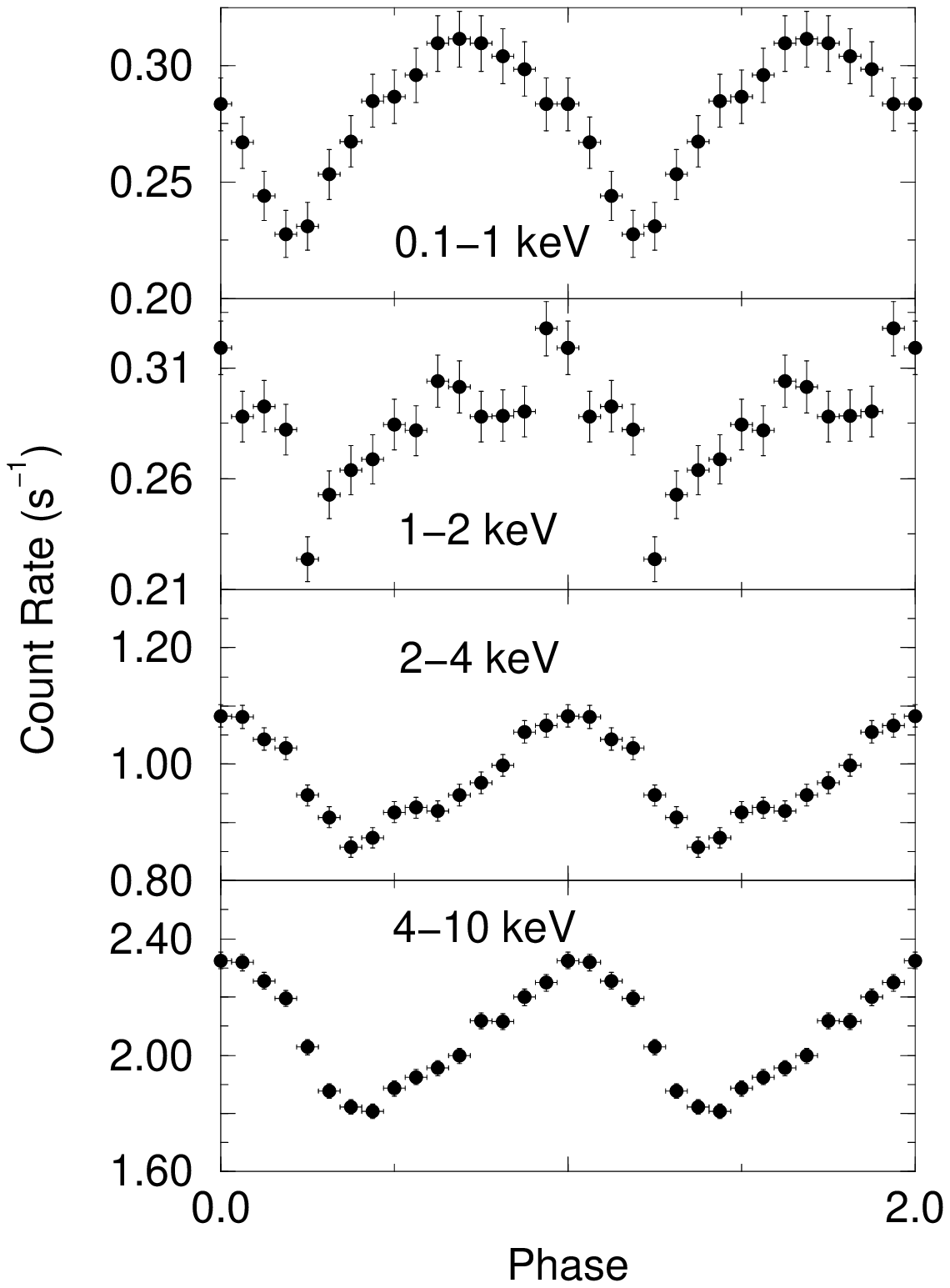}}}
\caption[]{Pulse profiles in 4 energy bands 
for the main- (left) and short-on (right) state
observations. The 0.1--1~keV and 1--2~keV profiles are from
the LECS and the 2--4~keV and 4--10~keV profiles from the
MECS. Pulse phase 0.0 is defined as the maximum of the 4--10~keV
folded lightcurves. The profiles are repeated for clarity}
\label{fig:pulse}
\end{figure*}

\subsection{Pulse period and profile}

The \src\ pulse periods were determined using only the non-dip
and non-eclipse data obtained during the main- and
short-on state observations. 
First, the arrival times of the photons were 
corrected
to the solar system barycenter. Then the arrival times were
additionally corrected to the \src\ center of mass using the ephemeris
of Deeter et al.\ (\cite{d:91}).  The periods were obtained with an
epoch-folding technique using the MECS data, while the (1$\sigma$)
uncertainties were determined by fitting the arrival times of sets of 6
averaged profiles. During the main- and short-on states the best-fit
periods are 1.2377268$\pm$0.0000009~s and 
1.2377272$\pm$0.00000015~s, respectively.

The pulse profiles obtained by folding the data with
these best-fit periods in different energy bands are shown in Fig.\
\ref{fig:pulse}. 
The well-known, strongly energy-dependent, pulse profile of
Her~X-1 is evident. In general, the main-on state profiles all exhibit
greater modulation depth than those obtained during the short-on state.
The smooth almost sinusoidal 0.1--1.0~keV profiles
are dominated by emission from the blackbody and Fe-L lines, while
the 2--4~keV and 4--10~keV profiles are dominated by the power-law.
The 2--4~keV main-on state profile is more sharply peaked and has
a larger amplitude than that
obtained during the short-on state. 
The 1--2~keV energy range clearly
marks the transition between the profiles dominated by these different
spectral components and it has a very structured appearance.
In this energy range the difference in
profiles is the largest between the two on-states.

\subsection{Pulse phase resolved spectrum}

The data obtained during intervals S1 and S2 (short-on) and M1 were
used to investigate the pulse phase dependence of the spectrum. 
Only the S1 and S2 intervals were included 
in the low-state accumulation since the spectra
obtained during S3 and S4 are clearly different from those obtained
elsewhere in the low-state. Both sets of data
were divided into 10 equal phase bins, 
which gives comparable (within a factor of two) statistics.  
(Note that this is half the number of
phase bins used by O97, since the total number of counts in both
observations is much smaller than that in the SVP observation.) The
spectra were rebinned and energy selected as above.
The 10 phase-resolved spectra were fit using the
``standard model'' except that \nh, the Fe-L line energies, the
Fe-K line energy
and width and LECS/MECS relative normalization were fixed at their
best-fit values obtained from the fits to the phase-averaged spectra
(Table~\ref{tab:ave_fits}). The best-fit values from the M1 and S1 fits
were used for the main- and short-on states phase resolved fits,
respectively.

\begin{figure*}
\centerline{\mbox{\includegraphics[width=8.5cm]{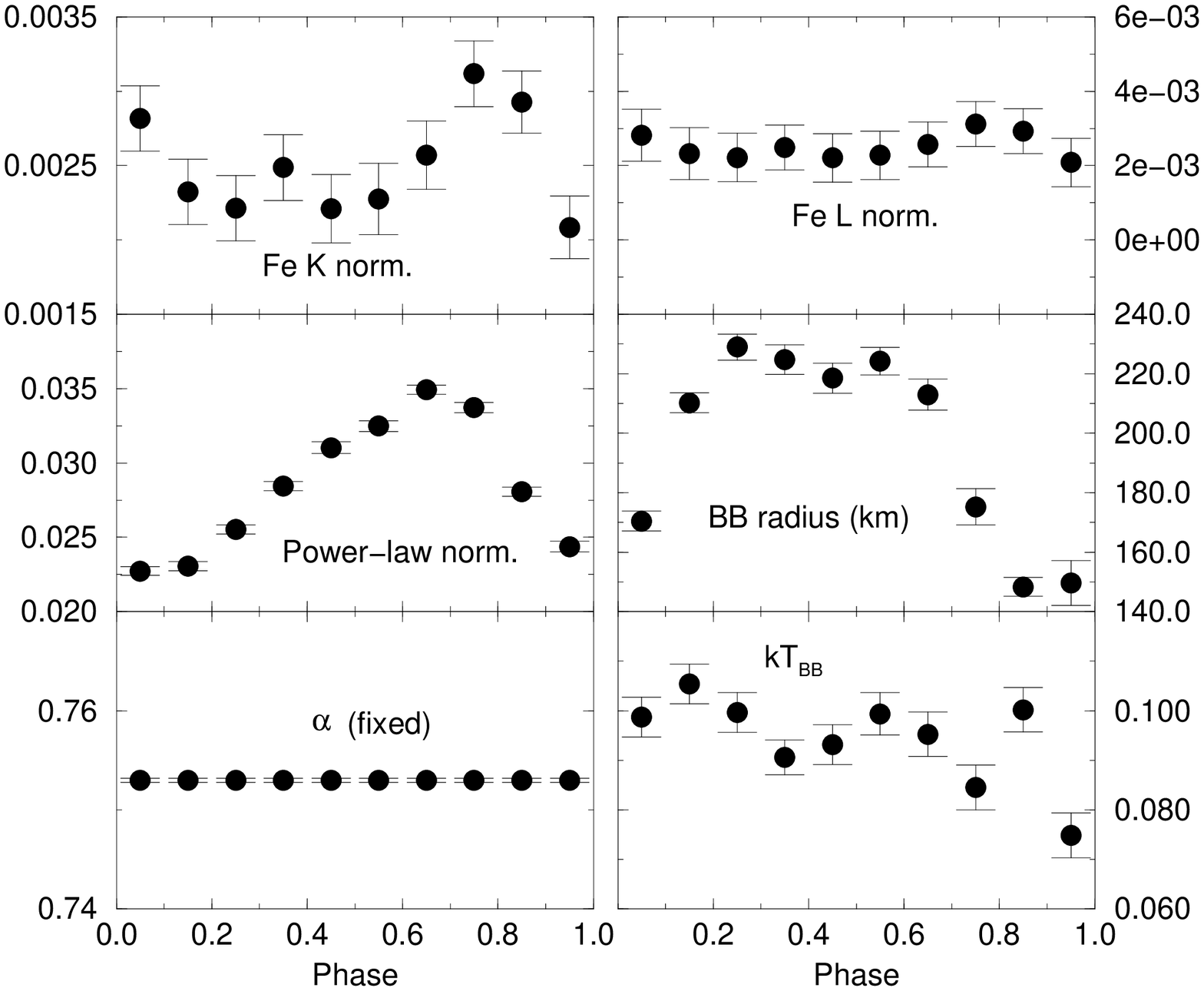}\hspace{1cm}\includegraphics[width=8.5cm]{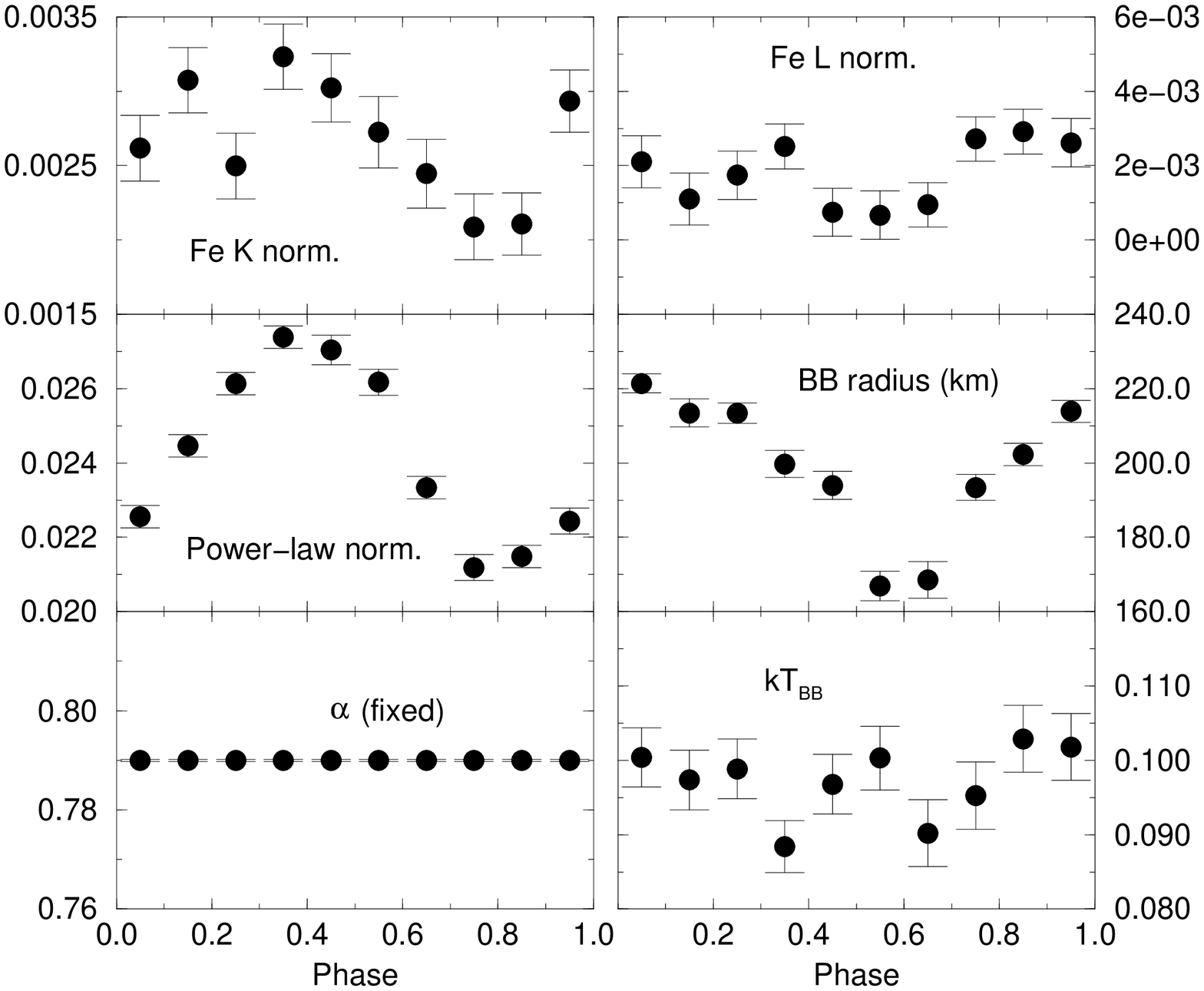}}}
\caption[]{Best-fit spectral parameters as a function of pulse phase
for the main- (left) and short-on (right) state observations.  
The units of normalization are: for the
power-law: Photons at 1 keV, cm$^{-2}$~keV$^{-1}$~s$^{-1}$, 
for the Gaussian features:
Photons cm$^{-2}$~s$^{-1}$, and for the blackbody:
L$_{39}$/d$_{10}^{2}$, where L$_{39}$ is the source luminosity in
10$^{39}$ erg s$^{-1}$ and d$_{10}^{2}$ is the source distance in
units of 10 kpc. Note that the power-law index, $\alpha$, was
kept constant. Uncertainties denote the 68\% confidence interval for 1
parameter of interest ($\Delta\chi^{2}=1.0$). Phases are arbitrary,
but identical for each observation}
\label{fig:phase}
\end{figure*}

The fits to the phase-resolved spectra give acceptable
values of \rchisq\ of typically 0.9--1.3 for 80--110 dof.
Fig.~\ref{fig:phase} shows the variations in best-fit spectral
parameters as a function of pulse phase.  Variations in all the fit
parameters, except the strengths of both Fe lines and the blackbody kT
are evident.  
Comparison of the left- and right-hand 
panels of Fig.\ \ref{fig:phase} shows that
the phase differences between the blackbody and power-law
normalizations are approximately the same for the main- and 
short-on states. The amplitude
of variation of the power-law component is smaller during the
short-on state observation ($\sim$25\%), compared to during the main-on state
observation ($\sim50$\%). The intensity of the Fe-L line appears to
be relatively constant.
The phase dependence of the Fe-K line intensity is
not well determined, but is consistent with having the same 
behavior as the power-law normalization, or with having no variation with
pulse phase. This is in contrast to the 
{\it Ginga} results of Choi et al. (\cite{c:94}),
where the blackbody and Fe-K line normalizations appear to be
correlated.

The ratio of the blackbody to power-law fluxes
are, during the main-on state, 18\% and 15\% for the pulse
minima and maxima, respectively and 16\% and 20\% during the short-on
state. Similar values can be derived from O97. This
means that the reprocessing fraction is similar in all the
observations, implying that the amount of reprocessing does not depend
strongly on $\Phi _{35}$.  The best-fit equivalent blackbody radii are
consistent with being identical between the main-on and the beginning
(S1 and S2) of the short-on at $\sim$210~km, while in the latter
phases of the short-on it decreases to $<$100~km. This may indicate
that about 75\% the blackbody reprocessing site
is obscured during the later parts of the short-on
state.

A cross-correlation between the best-fit power-law and blackbody
normalizations obtained during the main- and short-on states reveals
that in both cases the phase difference is consistent with a 
separation of $250 \pm 20\degmark$. This value
is marginally consistent with the phase difference of
$215 \pm 20\degmark$ obtained earlier in the main-on state by O97. 
However, due to the broad, asymmetric profiles that are different
in each observation (see Fig~\ref{fig:phase} and O97) it is difficult
to reliably measure the pulse phase difference between the two 
components and so probe any changes of (relative) positions
of the emission and reprocessing regions.

\section{Discussion}

We report on a long (5.7~day duration) observation of most of the
short-on state of \src\ with the BeppoSAX LECS and MECS together with
a short observation during the declining phase of the preceding
main-on state. We find that the main-on and early short-on state
spectra can be well fit with the standard blackbody and power-law
continuum model, together with Fe-L and Fe-K emission features.
During the later phases of the short-on there is evidence for
intervals of strong low-energy absorption. These intervals repeat
each 1.7~day orbital cycle, becoming increasingly long and occurring
earlier in each orbital cycle as the 
short-on state progresses. This behavior appears to be consistent with the
predictions of the warped disk model of e.g., Schandl \& Meyer (\cite{s:94})
and Schandl (\cite{s:96}) where the end of the short-on state is
caused by the accretion disk moving into the line of sight to the 
neutron star. These results therefore support the idea that the
\src\ system contains a warped, precessing, accretion disk. 

The shape and intensity of the S4 spectrum is very similar to that
measured by Mihara et al. (\cite{m:91}) during the {\it low-state}
of \src\ by {\it Ginga}. They model the spectrum with 
``scattered'' and absorbed (\nh = $10^{24}$~atom~cm$^{-2}$)
components and a power-law index equivalent to that found in the
main-on state. In addition, their best-fit Fe-K line energy, FWHM
and EW of $6.53 \pm 0.08$~keV, $0.8 \pm 0.4$~keV, and $1.0 \pm
0.1$~keV, respectively are all strikingly similar to those reported
here for interval S4 (Table~\ref{tab:ave_fits}). These BeppoSAX
observations therefore support the idea, first proposed by
Mihara et al. (\cite{m:91}), that the low-state \src\ emission contains
at least two components, one of which is significantly absorbed.

The pulse profiles (broad and almost sinusoidal) in both the main-
and short-on states are consistent with those reported by Deeter at al.\
(\cite{d:98}) obtained with {\it Ginga} at similar $\Phi _{35}$. 
Deeter et al.\
(\cite{d:98}) show that the pulse shape during the main-on state changes
rapidly from a clearly peaked shape to a much broader, more
sinusoidal, shape during the final stages of the main-on state. The
best-fit pulse periods demonstrate that \src\ is continuing its recent
spin-up trend as shown in Bildsten et al.\ (\cite{bi:97}).

The pulse
phase difference of the maxima of the power-law and
blackbody during the short-
and main-on state observations is consistent with being the same,
and consistent with the results of O97, obtained earlier in the
main-on state. 
Such a symmetry is expected for a
reprocessing region located at, or near, the inner edge of the 
accretion disk since at
$\Phi _{35}= 0.0$ and 0.5 the same
behavior is predicted. This is because of the symmetrical 
disk shape expected
unless the region of the disk where the reprocessing occurs is strongly warped
(see e.g.\
Heemskerk \& van Paradijs \cite{h:89} for this
``symmetry-rule''). This somewhat strengthens the association of the
blackbody component with reprocessed emission originating from the
inner edge of the accretion disk. Ideally, it would be preferable to track the
phase difference between the components over the entire 35 day cycle. However
\src\ is only visible as a strong X-ray source for two intervals
separated by about half the 35 day cycle, giving rise to a similar phase
dependence of the components. This, combined with the large pulse profile
changes, make it difficult to reliably restrict the location of 
the reprocessing site.

\acknowledgements 
The BeppoSAX satellite is a joint Italian and Dutch programme.
T. Oosterbroek acknowledges an ESA Research Fellowship.
The quick-look RXTE ASM data were provided by the ASM/RXTE team.
We thank the staff of 
the BeppoSAX Science Data Center for assistance with these
observations.

{}
\end{document}